%% file: main.tex
\documentclass[sigconf]{acmart}
\AtBeginDocument{%
  \providecommand\BibTeX{{%
    \normalfont B\kern-0.5em{\scshape i\kern-0.25em b}\kern-0.8em\TeX}}}

\usepackage{soul}
\usepackage{graphicx}
\usepackage{booktabs}
\usepackage{float}
\usepackage{overpic}
\usepackage{float}  
\usepackage{subfloat}
\usepackage{stfloats} 
\usepackage[skip=0.5\baselineskip]{caption}
\usepackage{bm}
\usepackage{amsmath}
\usepackage{booktabs}
\usepackage{multirow}
\usepackage{multicol}
\usepackage{enumitem}
\usepackage{subcaption}
\usepackage{threeparttable}
\usepackage{xspace}
\usepackage{color}
\usepackage[normalem]{ulem}
\usepackage{algorithmicx,algorithm}
\usepackage{algpseudocode}
\usepackage{algorithm,algpseudocode}
\usepackage{marvosym}   

\makeatletter
\def\@ACM@checkaffil{
    \if@ACM@instpresent\else
    \ClassWarningNoLine{\@classname}{No institution present for an affiliation}%
    \fi
    \if@ACM@citypresent\else
    \ClassWarningNoLine{\@classname}{No city present for an affiliation}%
    \fi
    \if@ACM@countrypresent\else
        \ClassWarningNoLine{\@classname}{No country present for an affiliation}%
    \fi
}
\makeatother

\usepackage{xspace}
\newcommand{\etal}{\emph{et al.}\xspace}

\newcommand{\ie}{\emph{i.e.,}\xspace}

\copyrightyear{2023}
\acmYear{2023}
\setcopyright{acmlicensed}\acmConference[CIKM '23]{Proceedings of the 32nd ACM International Conference on Information and Knowledge Management}{October 21--25, 2023}{Birmingham, United Kingdom}
\acmBooktitle{Proceedings of the 32nd ACM International Conference on Information and Knowledge Management (CIKM '23), October 21--25, 2023, Birmingham, United Kingdom}
\acmPrice{15.00}
\acmDOI{10.1145/3583780.3615134}
\acmISBN{979-8-4007-0124-5/23/10}

\begin{document}
\begin{sloppypar}
\title{Diffusion Augmentation for Sequential Recommendation}

\author{Qidong Liu}
\affiliation{%
  \institution{Xi'an Jiaotong University \& City University of Hong Kong}
}
\email{liuqidong@stu.xjtu.edu.cn}

\author{Fan Yan}
\affiliation{%
  \institution{Huawei Noah's Ark Lab}
}
\email{yanfan6@huawei.com}

\author{Xiangyu Zhao \Letter}
\thanks{\Letter \ \text{Xiangyu Zhao and Ruiming Tang are the corresponding authors}}
\affiliation{%
  \institution{City University of Hong Kong}
}
\email{xianzhao@cityu.edu.hk}

\author{Zhaocheng Du}
\affiliation{%
  \institution{Huawei Noah's Ark Lab}
}
\email{zhaochengdu@huawei.com}

\author{Huifeng Guo}
\affiliation{%
  \institution{Huawei Noah's Ark Lab}
}
\email{huifeng.guo@huawei.com}

\author{Ruiming Tang \Letter}
\affiliation{%
  \institution{Huawei Noah's Ark Lab}
}
\email{tangruiming@huawei.com}

\author{Feng Tian}
\affiliation{%
  \institution{Xi'an Jiaotong University}
}
\email{fengtian@mail.xjtu.edu.cn}

\renewcommand{\shortauthors}{Qidong Liu \etal}

\begin{abstract}
      Sequential recommendation (SRS) has become the technical foundation in many applications recently, which aims to recommend the next item based on the user's historical interactions. However, sequential recommendation often faces the problem of data sparsity, which widely exists in recommender systems. Besides, most users only interact with a few items, but existing SRS models often underperform these users. Such a problem, named the long-tail user problem, is still to be resolved. Data augmentation is a distinct way to alleviate these two problems, but they often need fabricated training strategies or are hindered by poor-quality generated interactions. To address these problems, we propose a \textbf{Diffu}sion \textbf{A}ugmentation for \textbf{S}equential \textbf{R}ecommendation (\textbf{DiffuASR}) for a higher quality generation. The augmented dataset by DiffuASR can be used to train the sequential recommendation models directly, free from complex training procedures. To make the best of the generation ability of the diffusion model, we first propose a diffusion-based pseudo sequence generation framework to fill the gap between image and sequence generation. Then, a sequential U-Net is designed to adapt the diffusion noise prediction model U-Net to the discrete sequence generation task. At last, we develop two guide strategies to assimilate the preference between generated and origin sequences. 
      To validate the proposed DiffuASR, we conduct extensive experiments on three real-world datasets with three sequential recommendation models. The experimental results illustrate the effectiveness of DiffuASR. 
      As far as we know, DiffuASR is one pioneer that introduce the diffusion model to the recommendation.
      The implementation code is available online \footnote{\url{https://github.com/liuqidong07/DiffuASR}}\footnote{\url{https://gitee.com/mindspore/models/tree/master/research/recommend/DiffuASR}}. 
\end{abstract}

\ccsdesc[500]{Information systems~Recommender systems}

\keywords{Sequential Recommendation; Data Augmentation; Diffusion Model}


\maketitle

\input{1Introduction}
\input{2Preliminary}

\input{3Method}
\input{4Experiment}

\input{5RelatedWork}

\input{6Conclusion}

\input{8Acknowledge}

\bibliographystyle{ACM-Reference-Format}
\bibliography{main}

\appendix

\end{sloppypar}
\end{document}

%% file: 1Introduction.tex
\begin{figure}[t]
\centering
\includegraphics[width=0.9\linewidth]{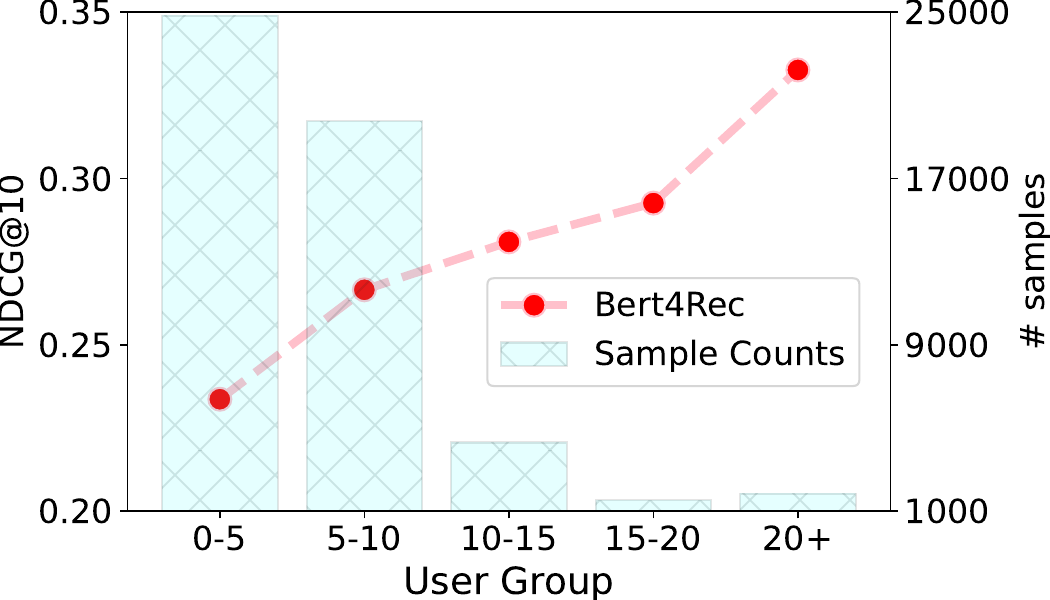}
\caption{The illustration for long-tail user problem.}
\label{fig:long_tail}
\vspace{-6mm}
\end{figure}

\section{Introduction} \label{sec:introduction}

The recommender system (RS) has shown promising developments in recent years. \textit{Sequential Recommendation} (SRS), as one important branch of RS, aims to recommend the next item for users based on their historical interactions~\cite{fang2020deep,wang2019sequential}. Thanks to the advancement of deep learning, SRS has been widely used in many applications, such as e-commerce~\cite{singer2022sequential,wang2020time} and smart education~\cite{zhang2019hierarchical}. The core of SRS is to capture the general pattern from the user's interaction sequences. Some recent works combine with transformer architecture to get such patterns and boost recommending performance, such as Bert4Rec~\cite{sun2019bert4rec} and SASRec~\cite{kang2018self}.

However, there exist two severe problems, which hinder the development of the sequential recommendation, \ie data sparsity and long-tail user problem. (i) \textbf{Data Sparsity Problem}. Sparsity is an inherent nature of real-world data, which means that only a few interactions occur among a large volume of users and items. It easily causes the sub-optimal problem for sequential recommendation~\cite{wang2021counterfactual}. (ii) \textbf{Long-tail User Problem}. This problem refers that users with few interactions often get worse recommendations. For illustration, we train a well-known SRS model, \ie Bert4rec~\cite{sun2019bert4rec}, on Beauty, a real-world dataset, and show the performance grouped by the user's historical interaction number. In Figure~\ref{fig:long_tail}, the histogram shows the number of users in each group and the line shows the recommending performance of each group. We can conclude that most users only own a few interactions, \ie interaction sequences shorter than 10, who are so-called long-tail users. However, existing SRS models often perform quite poorly for these users~\cite {liu2021augmenting,jiang2021sequential}, which makes most users unsatisfied with the recommender systems.

\begin{figure}[t]
\centering
\vspace{-1mm}
\includegraphics[width=0.81\linewidth]{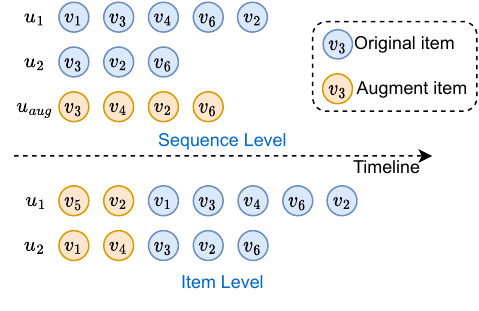}
\caption{Two types of data augmentation methods for SRS.}
\label{fig:aug_type}
\vspace{-6mm}
\end{figure}

Faced with the two problems, data augmentation is instinctive, because it can enrich the whole interaction records. We can categorize the existing data augmentation methods for SRS into two types, \ie sequence-level and item-level methods. (i) \textbf{Sequence-level Method}. As shown in Figure~\ref{fig:aug_type}, this line of methods~\cite{zhang2021causerec,wang2021counterfactual,wang2022learning,chen2022data} often generates new sequences or users to handle the problem of data sparsity. However, due to the noise contained in generated sequences, most sequence-level methods fabricate complex training strategies. For example, L2Aug~\cite{wang2022learning} adopts reinforcement learning to update SRS models and sequence generators iteratively. Wang \etal~\cite{wang2021counterfactual} propose to optimize the sequence generator based on a pre-trained SRS model and retrain the recommender on both original and augmented sequences. Compared with training models on augmented datasets directly, sequence-level methods are not convenient. Also, they are not tailored for long-tail users specifically. (ii) \textbf{Item-level Method}. As presented in Figure~\ref{fig:aug_type}, this category~\cite{liu2021augmenting,jiang2021sequential} generates pseudo pre-order items for the interaction sequence of each user. In general, the augmented dataset by this category can be used to train the SRS models directly. Besides, item-level methods can enhance the long-tail users, because they prolong the short interaction sequence. 
Nevertheless, ASReP~\cite{liu2021augmenting} and BiCAT~\cite{jiang2021sequential} focus on how to use the augmented dataset more efficiently but ignore the augmentation quality itself. They both train a SASRec~\cite{kang2018self} reversely as an item generator and get the augmented pre-order items iteratively. However, the top-1 accuracy of the existing SRS model is relatively low, so the iterative generation may cause the challenge of poor quality augmentation and distribution drift from the original sequence. These challenges block the benefits of data augmentation for data sparsity and long-tail user problems.

To handle the challenges mentioned above, we propose a \textbf{Diffu}sion \textbf{A}ugmentation for \textbf{S}equential \textbf{R}ecommendation (\textbf{DiffuASR}). The adoption of the diffusion model has two-fold merits. First, the diffusion model has been proven that it is powerful to generate high-quality contents, such as images~\cite{rombach2022high,saharia2022photorealistic} and texts~\cite{li2022diffusion,he2022diffusionbert,li2023diffusion}, so it has the potential to generate high-quality augmented items. Second, we adopt the diffusion model to generate the sequence in one run, which can guarantee that the whole augmented sequence has a more similar distribution to the original one.
However, some problems emerge while combining the diffusion model with the sequential recommendation. First, the diffusion model is initially designed to generate an image, a continuous-valued matrix. By comparison, the interaction history is a 1-dimensional discrete-valued sequence, which is hard to be generated directly from the diffusion model. 
To fill the gap between image and sequence generation, we derive a diffusion-based pseudo sequence generation framework for augmenting sequential recommendation. 
Then, U-Net~\cite{ronneberger2015u} is an essential component for the general diffusion model, but it is difficult to capture the sequential information of the embedding sequence. Therefore, we design an SU-Net compatible with the embedding sequence. 
Besides, the augmented pseudo items should be relevant to the preference of the user's original intent, because not relevant items may cause noise for training~\cite{zhang2022hierarchical,chen2022denoising}. Inspired by the guided diffusion model~\cite{dhariwal2021diffusion,ho2022classifier}, we design a classifier-guide and a classifier-free strategy for DiffuASR to guide the generation process.
The contributions of this paper are concluded as follows:

\begin{itemize}[leftmargin=*]
    \item We design a novel data augmentation framework based on diffusion model for the sequential recommendation. This framework fills the gap between discrete-valued item identity and generated real-valued matrix, which is one of the pioneer works to combine the diffusion model with the recommender system;
    
    \item To fit the U-Net for the sequence generation task, we devise an SU-Net that can recover the embedding sequence matrix. Also, to guide the diffusion model to generate the items that are more corresponding to the user's intent, we design a classifier-guide and a classifier-free strategy for DiffuASR;
    
    \item We conduct comprehensive experiments on three real-world datasets to illustrate the effectiveness of our augmentation method. Besides, the experimental results on three different sequential recommendation models show the generality of the DiffuASR.

\end{itemize}

%% file: 2Preliminary.tex
\section{Preliminary} \label{sec:preliminary}

In this section, we will introduce the problem definition and notations used in this paper briefly. Besides, some fundamentals of the diffusion model are presented.

\subsection{Problem Definition and Notations}

The sequential recommendation aims to recommend the next item for users according to their historical interactions, so the records of one user are often arranged by timeline. If we let $\mathcal{U}=\{u_1,u_2,...,u_{|\mathcal{U}|}\}$ and $\mathcal{V}=\{v_1,v_2,...,v_{|\mathcal{V}|}\}$ denote the set of users and items, respectively, the interaction history of one user $u$ can be written as $\mathcal{S}_u=\{v_1^{(u)},...,v_k^{(u)},....,v_{n_u}^{(u)}\}$. $n_u$ is the number of users' historical interactions. We omit the sub-script $(u)$ in the following sections for simplicity. Then, the problem of sequential recommendation can be presented as: \textit{Given a user's interaction history $\mathcal{S}_u$, predict the most possibly interacted item at step $n_u+1$, \ie $v_{n_u+1}$}. It is formulated as follows:
\begin{equation} \label{eq1}
    arg \max_{v_i \in \mathcal{V}} P(v_{n_u+1}=v_i|\mathcal{S}_u)
\end{equation}

Item-level augmentation methods generate the pre-order items for the raw sequence. For example, given a historical sequence $\mathcal{S}_{raw}=\{v_1,...,v_k,....,v_{n_u}\}$, the augmentation items are inserted before the $v_1$. Let $\mathcal{S}_{aug}=\{v_{-M},v_{-M+1},...,v_{-1}\}$ denote the augmented sequence for $\mathcal{S}_{raw}$, where $M$ is the number of augmented items. Then, the original sequence is transformed into $\mathcal{S}'=\{v_{-M},...,v_{-1},v_1,...., v_{n_u}\}$ by augmentation. As illustrated in Section~\ref{sec:introduction}, the goal in this paper is to generate high-quality $S_{aug}$.

\begin{figure*}[t]
\centering
\includegraphics[width=0.9\linewidth]{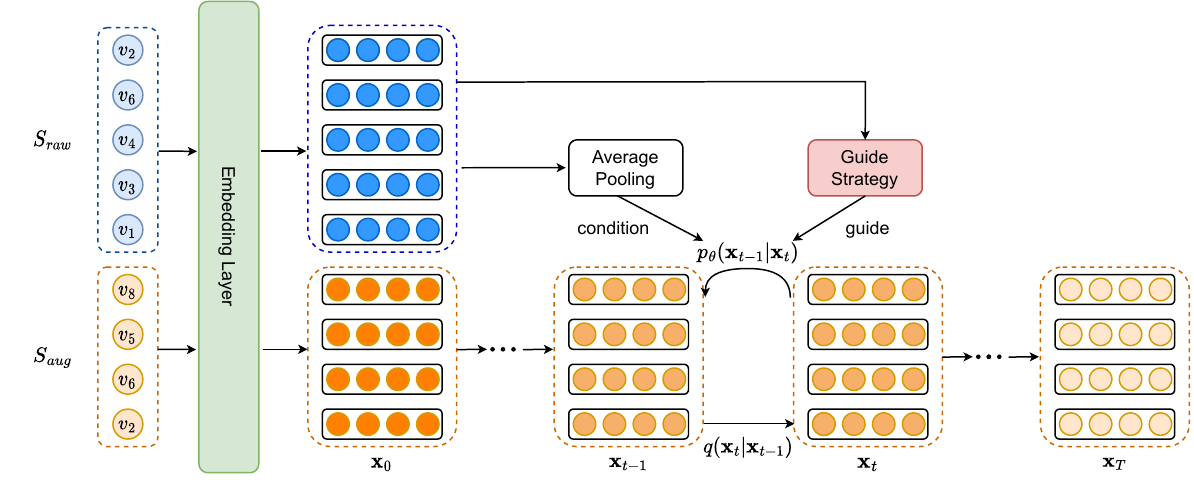}
\caption{The overview of the proposed Diffusion Augmentation for Sequential Recommendation (DiffuASR).}
\label{fig:overview}
\vspace{-4mm}
\end{figure*}

\subsection{Diffusion Model}

Diffusion model~\cite{ho2020denoising} has become prevalent in recent years, due to its strong capability in various generation tasks, such as images~\cite{rombach2022high,saharia2022photorealistic} and texts~\cite{li2022diffusion,he2022diffusionbert,li2023diffusion}. For an easier understanding of the proposed method, we will briefly introduce the fundamental knowledge of the diffusion model before presenting DiffuASR. In general, the diffusion model is composed of two processes, \ie forward process and the reverse process. The forward process is gradually adding noise to original data samples, which perturbs samples approaching a Gaussian noise. On the contrary, the reverse process recovers the original sample step by step via predicting the added noise. Then, we will detail these two processes.

\textbf{Forward Process}. Let $\mathbf{x}_0 \sim q(\mathbf{x}_0)$ denotes the original data samples. The forward process is a Markov Chain that gradually adds Gaussian noise to $\mathbf{x}_0$, which can be formulated as follows:
\begin{equation} \label{eq2}
    \begin{aligned}
        q(\mathbf{x}_{1:T}|\mathbf{x}_0) &:=\prod_{t=1}^T q(\mathbf{x}_t|\mathbf{x}_{t-1}) \\
        q(\mathbf{x}_t|\mathbf{x}_{t-1}) &:= \mathcal{N}(x_t;\sqrt{1-\beta_t}\mathbf{x}_{t-1}, \beta_t\mathbf{I})
    \end{aligned}
\end{equation}
\noindent where $T$ is the number of steps to add noise and $\beta_1,...,\beta_T$ are the variance schedule. Since all data samples are perturbed to normal Gaussian noise, the diffusion step number $T$ and variance schedule $\{\beta_t\}^T_{t=1}$ are adjustable to control how much noise is added at each diffusion step. However, it is difficult to learn the variance $\beta_t$ directly, so a reparameterization trick~\cite{kingma2013auto} is adopted here and thus transforms the forward process as follows:
\begin{equation} \label{eq3}
    q(\mathbf{x}_t|\mathbf{x}_0)=\mathcal{N}(\mathbf{x}_t;\sqrt{\bar{\alpha}_t \mathbf{x}_t},(1-\bar{\alpha}_t) \mathbf{I})
\end{equation}
\noindent where $\alpha_t=1-\beta_t$ and $\bar{\alpha}_t=\prod^t_{s=1}\alpha_s$.

\textbf{Reverse Process}. The reverse process aims to recover the original data sample $\mathbf{x}_0$ from the completely perturbed $\mathbf{x}_T \sim \mathcal{N}(0,\mathbf{I})$. We can formulate the reverse process by the Markov chain as:
\begin{equation}    \label{eq4}
    \begin{aligned}
        p_{\theta}(\mathbf{x}_{0:T}) &= p(\mathbf{x}_T) \prod_{t=1}^T p_{\theta}(\mathbf{x}_{t-1}|\mathbf{x}_t) \\
        p_{\theta}(\mathbf{x}_{t-1}|\mathbf{x}_t) &= \mathcal{N}(\mathbf{x}_t;\mu_{\theta}(\mathbf{x}_t,t),\Sigma_{\theta}(\mathbf{x}_t,t))
    \end{aligned}
\end{equation}
\noindent where $\mu_{\theta}(x_t,t)$ and $\Sigma_{\theta}(x_t,t)$ are the mean and variance that are parameterized by $\theta$. Based on the Gaussian distribution parameterization~\cite{ho2020denoising}, we can set $\Sigma_{\theta}(\mathbf{x}_t,t)=\sigma_t^2 \mathbf{I}$ as constants, where $\sigma_t^2=\frac{1-\bar{\alpha}_{t-1}}{1-\bar{\alpha}_t} \beta_t$. As for the mean of the distribution, it can be written as:
\begin{equation}    \label{eq5}
    \mu_{\theta}(\mathbf{x}_t,t)=\frac{1}{\sqrt{\alpha_t}}(\mathbf{x}_t-\frac{\beta_t}{\sqrt{1-\bar{\alpha}_t}})\epsilon_{\theta}(\mathbf{x}_t,t)
\end{equation}
Therefore, the diffusion models are mainly parameterized by the $\epsilon_{\theta}$, which is usually an U-Net~\cite{ronneberger2015u} based architecture.

\textbf{Optimization}: The objective of the general diffusion model is to pull the posterior distribution in the forward process and prior distribution in the reverse process closer, so the objective function can be written as the KL divergence form. For simplification, it can also be formulated as~\cite{ho2020denoising}:
\begin{equation}    \label{eq6}
    L_{simple}=\mathbb{E}_{t,\mathbf{x}_0,\epsilon}[\|\epsilon-\epsilon_{\theta}(\sqrt{\bar{\alpha}_t}\mathbf{x}_0+\sqrt{1-\bar{\alpha}_t \epsilon,t})\|^2]
\end{equation}

%% file: 3Method.tex
\section{DiffuASR}  \label{sec:method}

In this section, we will detail the design of our Diffusion Augmentation for Sequential Recommendation (DiffuASR). As a start, we present the overview of the augmentation framework in Section~\ref{subsec:overview}. In Section~\ref{subsec:sunet}, we will refer to the proposed SU-Net architecture used to recover the embedding sequence. Next, we introduce how to guide the diffusion model to generate pseudo items corresponding to the raw sequence in Section~\ref{subsec:guide}. Finally, the train and augment process are illustrated in Section~\ref{subsec:opt}.

\begin{figure*}[t]
\centering
\includegraphics[width=0.9\linewidth]{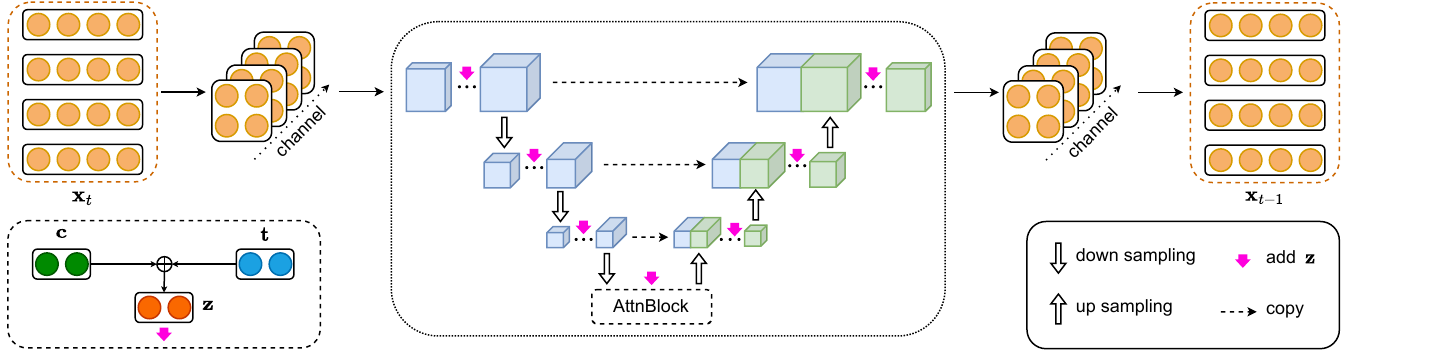}
\caption{The architecture of the proposed SU-Net.}
\label{fig:sunet}
\vspace{-5mm}
\end{figure*}

\subsection{Framework} \label{subsec:overview}
As Figure~\ref{fig:overview} shows, we design the DiffuASR framework for data augmentation. The augmented sequence $S_{aug}$ is the generative objective of the diffusion model, while the $S_{raw}$ is the raw sequence of the user.
The task of data augmentation for sequential recommendation is to generate a 1-dimensional discrete-valued sequence, while general diffusion models produce a 2-dimensional continuous-valued image matrix. To fill this gap, we derive the embedding procedure for the forward process and the rounding procedure for the reverse process. Then, the augmented sequence $S_{aug}$ should have a similar distribution of $S_{raw}$ to avoid noisy training~\cite{zhang2022hierarchical,chen2022denoising}, so we devise the guide procedure for reverse process to assimilate them. In the rest of this section, we will detail the DiffuASR framework according to the forward process, reverse process and guide procedure.

    \subsubsection{\textbf{Forward Process}}
    
        As illustrated in Section~\ref{sec:preliminary}, the augmented sequence is a series of item identities, which can be written as $S_{aug}=\{v_{-M},v_{-M+1},...,v_{-1}\}$. To get the representation of each item, we adopt a learnable embedding table to transform the item identity $v_i$ into dense representation $\mathbf{e}_i$. Let $\mathbf{E} \in \mathbb{R}^{|\mathcal{V}| \times d}$ denotes the embedding table, where $d$ is the dimension of embedding. After the embedding table, we can get the stacked item embedding (an embedding sequence), denoted as $\mathbf{x}_0=\{\mathbf{e}_{-M},\mathbf{e}_{-M+1},...,\mathbf{e}_{-1}\}$. Here, the 1-dimensional discrete-valued sequence is converted to a 2-dimensional continuous-valued matrix, similar to the input type of the general diffusion model. Then, following the Equation~\eqref{eq2}, we add the Gaussian noise to the embedding sequence $\mathbf{x}_t$ iteratively and get the $\mathbf{x}_T \sim \mathcal{N}(0,\mathbf{I}^{M \times d})$. During the forward process, except for the diffusion step $T$, the variance schedule is also important for controlling the noise intensity in each step. The linear schedule is adopted in DiffuASR due to better performance.

    \subsubsection{\textbf{Reverse Process}}
    
        The task of the reverse process is to recover the embedding sequence $\mathbf{x}_0$ from $\mathbf{x}_T$. According to Equation~\eqref{eq4}, this process removes the Gaussian noise step by step, so the core of the reverse process is to predict the added noise for each diffusion step. After simplification, the noise can be predicted by the $\epsilon_{\theta}(\mathbf{x}_t,t)$ as the Equation~\eqref{eq5} shows. In general, $\epsilon_{\theta}(\mathbf{x}_t,t)$ is implemented by U-Net~\cite{ronneberger2015u}. However, U-Net often absorbs the image matrix instead of sequence embedding, which contains different meanings along the column and row dimensions. Therefore, we propose a sequential U-Net (SU-Net) for $\epsilon_{\theta}$, which will be referred to in Section~\ref{subsec:sunet}. Finally, the predicted $p_{\theta}(x_{t-1}|x_t)$ helps recover the augmented embedding sequence $\hat{\mathbf{x}}_t=\{\hat{\mathbf{e}}_{-M},\hat{\mathbf{e}}_{-M+1},...,\hat{\mathbf{e}}_{-1}\}$. To get the augmented sequence, we design a rounding procedure. This procedure uses the cosine similarity function to decide the augmented item for each position $j \in [-M,-1]$:
        \begin{equation} \label{eq7}
            v_j=\max_{v_i \in \mathcal{V}} \ {\rm sim}(\hat{\mathbf{e}}_j, \mathbf{e}_i)
        \end{equation}
        
        \noindent where $\mathbf{e}_i$ is the embedding of item $v_i \in \mathcal{V}$ and ${\rm sim}(\cdot, \cdot)$ represents the cosine similarity function. The reason for using cosine similarity here is that the generated embedding has a different embedding length compared with the original item embedding. 

    \subsubsection{\textbf{Guide Procedure}}
    
        Only forward and reverse processes cannot guarantee that the generated contents are desired, because no control signal is injected while generating. Thus, a guide procedure is needed. Specifically, we hope the generated items can correspond to the preference of users. Users' preference is often contained in their historical interactions, so we utilize $S_{raw}$ to guide the generation process. The guide procedure is in two folds. On the one hand, we add the information of $S_{raw}$ to $\epsilon_{\theta}$, because the predicted noise controls the direction of generation. In detail, the embedding sequence is converted to conditional vector $\mathbf{c}=\rm{Avg}(e_1,e_2,...,e_{n_u})$ and injected to $\epsilon_{\theta}(\mathbf{x_t},t,\mathbf{c})$. On the other hand, we design two guide strategies for further guidance, which will be detailed in Section~\ref{subsec:guide}.

\subsection{SU-Net} \label{subsec:sunet}

In this section, we will introduce the proposed Sequential U-Net (SU-Net). U-Net is widely used in the diffusion model, but it is devised for images and may lose the sequence information if we use it to handle the embedding sequence directly. Therefore, we propose altering the U-Net input to fit the embedding sequence. 

The architecture of SU-Net is shown in Figure~\ref{fig:sunet}. To capture the sequential information of $\mathbf{x}_t$, we set the sequence dimension as the channel of the image, because the channel is modeled independently in the convolutional networks. However, the input of the image matrix becomes 1-dimensional, \ie only the embedding dimension. For full usage of convolutional neural networks, we reshape each 1-dimensional embedding into a matrix with the shape of $(\sqrt{d},\sqrt{d})$. For example, if the size of the embedding vector is 64, the embedding can be converted to a $8 \times 8$ matrix. As a result, the input embedding sequence is transformed into an $M$ channels matrix with $\sqrt{d}$ length and width, which will be input into the main architecture of SU-Net.

The main architecture of SU-Net is similar to the U-Net~\cite{ronneberger2015u}. It contains several steps of down-sampling and up-sampling. Before the down-sampling, multiple resnet blocks with convolutional neural networks are imposed on the multi-channel matrix. Then, a middle AttnBlock equipped with a resnet and attention layer is used to remove the redundant information. During the up-sampling process, the matrix from down-sampling is copied and concatenated to fuse more features. The output matrix has the same size as the input, so we reshape each $(\sqrt{d},\sqrt{d})$ matrix back to the 1-dimensional embedding vector to get the $\mathbf{x}_{t-1}$.

The diffusion step $t$ and conditional vector $\mathbf{c}$ are also vital for $\epsilon_{\theta}(\mathbf{x}_t,t,\mathbf{c})$ to learn the noise. To represent the diffusion step, we convert the scalar $t$ into a vector $\mathbf{t}$ by sinusoidal embedding technique~\cite{vaswani2017attention}, which has the same size of $\mathbf{c}$. Then, we add the step vector and conditional vector, \ie $\mathbf{z}=\mathbf{c}+\mathbf{t}$. Finally, $\mathbf{z}$ is added to each resnet block of SU-Net architecture for the injection of these two types of information. Specifically, $\mathbf{z}$ is transformed to the same size of the input matrix by linear layer and added to the input matrix before the convolution layer. Thus, the diffusion step and conditional vector can control the $\epsilon_{\theta}$ to predict the noise.

\subsection{Guide Strategy} \label{subsec:guide}

High-quality augmentation should correspond to the preference contained in the raw user's interaction sequence, because unrelated items can be considered as noisy interactions and thus harm the sequential recommendation model~\cite{zhang2022hierarchical,chen2022denoising}. To make the augmented items more relevant to the original sequences, we design two guide strategies at the reverse process for DiffuASR, \ie classifier guide and classifier free. Then, we will detail these two strategies. 

\subsubsection{\textbf{Classifier Guide Strategy}}

    In the reverse process, each step updates towards the gradient direction of raw data distribution. For this reason, Dhariwal \etal~\cite{dhariwal2021diffusion} propose to add a gradient of image classifier to $\epsilon$ to guide the generation towards a specific class, so the noise prediction can be reformulated as follows:
    \begin{equation} \label{eq8}
        \hat{\epsilon}=\epsilon_{\theta}(\mathbf{x}_t)-\gamma \cdot \sqrt{1-\bar{\alpha}}_t \nabla_{\mathbf{x}_t}log \ p_{\phi}(y|\mathbf{x}_t)
    \end{equation}
    \noindent where $\gamma$ is the scale for guidance and $p_{\phi}(\cdot)$ represents the classifier. However, the recommendation task is different from the classification task, which makes the original classifier unsuitable. The augmented sequence $S_{aug}$ is the pre-order sequence of the first item in $S_{raw}$, so $v_1$ can be regarded as the next recommendation item of $S_{aug}$. Inspired by this thought, we propose the sequential recommendation model as the ``classifier'' and the Equation~\eqref{eq8} in DiffuASR can be written as follows:
    \begin{equation} \label{eq9}
        \hat{\epsilon}=\epsilon_{\theta}(\mathbf{x}_t,t,\mathbf{c})-\gamma \cdot \sqrt{1-\bar{\alpha}}_t \nabla_{\mathbf{x}_t}log \ p_{\phi}(v_1|S_{aug})
    \end{equation}
    \noindent where $\phi$ is a pre-trained sequential recommendation model and $p(\cdot)$ is binary cross-entropy loss instead of cross-entropy loss in~\cite{dhariwal2021diffusion}. 

\subsubsection{\textbf{Classifier Free Strategy}}

    However, the classifier guide strategy needs a pre-trained ``classifier'', which costs much time and computation. 
    To free from the laborious pre-training procedure, we propose to utilize the classifier free strategy~\cite{ho2022classifier} as the other variant of our DiffuASR. The predicted noise of classifier free strategy in the reverse process is formulated as follows:
    \begin{equation} \label{eq10}
        \hat{\epsilon} = (1+\gamma) \cdot \epsilon_{\theta}(\mathbf{x}_t,t,\mathbf{c})-\gamma \cdot \epsilon_{\theta}(\mathbf{x}_t,t,\mathbf{e}_{padding})
    \end{equation}
    \noindent where $\gamma$ is the scale for guidance and $\mathbf{e}_{padding}$ is the padding vector.

\let\oldnl\nl
\newcommand{\nonl}{\renewcommand{\nl}{\let\nl\oldnl}}
\begin{algorithm}[t]
\caption{Train and Augment process of DiffuASR} \label{alg:train}
\raggedright

\begin{algorithmic} [1]
    \State Indicate the augmented number $M$ for each sequence.
    \State Indicate the variance schedule and calculate $\{\beta_t\}_{t=1}^T$.
    \State Indicate the guidance scale $\gamma$ and guide strategy. 
    \State Pre-train a sequential recommendation model $\phi$.
    \State Select out the sequence with $n_u>M$ from $\mathcal{D}$ to form $\mathcal{D}_T$.
\end{algorithmic}

\textbf{Train Process} 
\setcounter{algorithm}{4}
\begin{algorithmic} [1]
    \makeatletter
    \setcounter{ALG@line}{5}
    \For {$S$ in $\mathcal{D}_T$}
        \State Take out the first $M$ items as $S_{aug}$ and the rest as $S_{raw}$.
        \State Conduct the forward process by Equation~\eqref{eq3}.
        \If {Guide Strategy is Classifier Guide}
            \State Calculate the $\hat{\epsilon}$ by Equation~\eqref{eq9}.
        \Else
            \State Calculate the $\hat{\epsilon}$ by Equation~\eqref{eq10}.
        \EndIf
        \State Conduct the reverse process by Equation~\eqref{eq4}.
        \State Calculate the loss by Equation~\eqref{eq6} and update the parameters $\theta$ by gradient descent.
    \EndFor
\end{algorithmic}

\textbf{Augment Process}
\setcounter{algorithm}{15}
\begin{algorithmic} [1]
    \makeatletter
    \setcounter{ALG@line}{16}
    \State Set the augmented dataset $\mathcal{D}_A$
    \For {$S$ in $\mathcal{D}$}
        \If {Guide Strategy is Classifier Guide}
            \State Calculate the $\hat{\epsilon}$ by Equation~\eqref{eq9}.
        \Else
            \State Calculate the $\hat{\epsilon}$ by Equation~\eqref{eq10}.
        \EndIf
        \State Conduct the reverse process by Equation~\eqref{eq4} based on $S$ and $\epsilon$.
        \State Get the augment sequence $\hat{S}_{aug}$ and add $[\hat{S}_{aug},S]$ to $\mathcal{D}_A$.
    \EndFor
    \State Get the augmented dataset $\mathcal{D}_A$.

\end{algorithmic}
\end{algorithm}

\subsection{Train and Augment} \label{subsec:opt}

In this section, we will introduce the train and augment process of DiffuASR. We conclude the process in Algorithm~\ref{alg:train} to be more readable. At first, we need to indicate some hyper-parameters in advance (lines 1-3), such as augmented number $M$ and guidance scale $\gamma$. Then, we make up the dataset $\mathcal{D}_T$ for training DiffuASR (line 5). Since we do not have the ground-truth augment sequence, the pre-order items in the raw sequence are regarded as augmentation for training. Thus, only the sequences with a length longer than the augmented number $M$ can be used for training DiffuASR. Then, we take out the first $M$ items in a sequence for diffusion (line 7). Next, the forward and reverse processes are conducted and the parameters $\theta$ of SU-Net are updated (lines 8-15). When the model gets convergence, the train process ends. As for the augment process, the guided reverse process is conducted based on well-trained $\epsilon_{\theta}$ (line 19-24) to generate pre-order sequence $\hat{S}_{aug}$. Then, $\hat{S}_{aug}$ is concatenated before the raw sequence, which forms the augmented sequence. After augmentation for each raw sequence in $\mathcal{D}$, we can get the augmented dataset $\mathcal{D}_A$. Unlike the sequence-level augmentation methods, $\mathcal{D}_A$ can be directly used for training sequential recommendation models to enhance their performance.

%% file: 4Experiment.tex
\section{Experiment}

In this section, we will introduce the extensive experiments conducted on three real-world datasets. We analyze the experimental results to illustrate the following research problems (\textbf{RQ}):
\begin{itemize}[leftmargin=*]
    \item \textbf{RQ1:} How DiffuASR performs compared with other item-level augmentation methods? Is DiffuASR a general augmentation method for various sequential recommendation models?
    \item \textbf{RQ2:} Can DiffuASR alleviate the long-tail user problem?
    \item \textbf{RQ3:} How does the number of augmented items affect the recommending performance of DiffuASR?
    \item \textbf{RQ4:} How do the hyper-parameters of the diffusion model affect the performance of DiffuASR?
\end{itemize}

\begin{table}[]
\centering
\caption{The Statistics of Datasets}
\begin{tabular}{ccccc}
\toprule[1.5pt]
Dataset & \# Users & \# Items & Sparsity & Avg.length \\ 
\midrule
\midrule
Yelp & 30,431 & 20,033 & 99.95\% & 10.40 \\
Beauty & 50,498 & 57,289 & 99.99\% & 7.76 \\
Steam & 332,925 & 13,047 & 99.91\% & 11.06 \\ 
\bottomrule[1.5pt]
\end{tabular}
\label{tab:dataset}
\vspace{-4mm}
\end{table}

\subsection{Experiment Settings}

\subsubsection{\textbf{Dataset}}

    We conduct comprehensive experiments on three representative real-world datasets, \ie Yelp, Amazon Beauty and Steam. \textbf{Yelp}\footnote{\url{https://www.yelp.com/dataset}} is a business recommendation dataset, which records the check-in history of users. We only use the interactions that occurs after Jan 1st, 2019, since the whole dataset is too large. \textbf{Beauty}\footnote{\url{https://cseweb.ucsd.edu/~jmcauley/datasets.html\#amazon_reviews}} is an review dataset on an e-commerce platform. Amazon is a large-scale product review dataset published in~\cite{mcauley2015image} and Beauty is one sub-category. \textbf{Steam}\footnote{\url{https://cseweb.ucsd.edu/~jmcauley/datasets.html\#steam_data}} dataset collects the players' reviews for video games on the Steam platform. This dataset is published in~\cite{kang2018self}.
    
    We preprocess all the datasets referring to~\cite{kang2018self,sun2019bert4rec}. For sequential recommendation task, we regard all reviewed or rated records as implicit interactions. Then, to fulfill the test split requirement and investigate the long-tail users, we only remove the users with fewer than three interactions. As for data split, we take the last item $v_{n_u}$ as the test and the penultimate item $v_{n_u-1}$ as the validation. The rest of the raw sequence is split into the training data. After the preprocessing, the statistics of the datasets are shown in Table~\ref{tab:dataset}.

\subsubsection{\textbf{Baselines}}

    To show the effectiveness of the proposed augmentation method, we compare DiffuASR with some item-level baseline methods:
    \begin{itemize}[leftmargin=*]
        \item \textbf{None}. No augmentation for this baseline. The recommendation models are trained directly on the original dataset $\mathcal{D}$. 
        \item \textbf{Random}. Augment each sequence by randomly selecting items from the whole item set, \ie $v_{aug} \in \mathcal{V}$.
        \item \textbf{Random-Seq}. Select items from the original sequence randomly as the augmented items, \ie $v_{aug} \in \mathcal{S}_u$.
        \item \textbf{ASReP}~\cite{liu2021augmenting}. 
        This baseline train a SASRec~\cite{kang2018self} model in a reverse pattern and generate the augmentation item iteratively. The pretrain-finetune scheme used in this baseline can only be compatible with SASRec. In contrast, we want to test the generality for several SRS models and focus on the generation quality. Thus, we only use the reverse SASRec as the augmentor and do not adopt extra training tricks. 
        The BiCAT~\cite{jiang2021sequential} uses the same augmentation method, so we do not discuss about it further.
    \end{itemize}

\subsubsection{\textbf{Backbone Recommendation Models}}

    Generality is a vital characteristic of the data augmentation method. To illustrate whether the proposed DiffuASR possesses such characteristic, we combine competing augmentation methods and our DiffuASR with several sequential recommendation models:
    \begin{itemize}[leftmargin=*]
        \item \textbf{Bert4Rec~\cite{sun2019bert4rec}}. It is a sequential recommendation model equipped with bi-directional self-attention layers. Besides, it adopts the cloze task to train the model, which is different from general sequential recommendation models.
        \item \textbf{SASRec~\cite{kang2018self}}. Different from Bert4Rec, SASRec proposes a causal self-attention layer to capture the transitional patterns of the user's interaction sequence.
        \item \textbf{S3Rec~\cite{zhou2020s3}}. S3Rec designs four self-supervised tasks to embed the information of item attributes and sequence correlation into the model. Then, it finetunes the model using the next item prediction task. Since the referred augmentation methods do not consider the item attribute, we only implement masked item prediction and segment prediction tasks for pre-train.
    \end{itemize}

\subsubsection{\textbf{Implementation Details}}

    We choose the best hyper-parameters by the HR@10 metric on the validation set. For DiffuASR, we tune the augment number $M$ from $\{4,6,8,10,12,14,16\}$ and the guidance scale $\gamma$ from $\{0.1,1,10,100\}$. The number of diffusion steps is fixed at 1,000. We set the dimension of item embedding as 64. After augmentation, $\mathcal{D}_A$ is used to train the sequential recommendation model directly. As for the sequential recommendation model, we fix the maximum sequence length as 200 and the dropout rate as 0.6 for all three datasets. Other hyper-parameters of the three SRS models follow the original paper. For Bert4Rec, the cloze task is used for training. For SASRec and finetune stage of S3Rec, sequence-to-one training scheme is adopted. To optimize the DiffuASR and sequential recommendation models, we apply Adam optimizer~\cite{kingma2014adam}, and set the batch size and learning rate as 512 and 0.001, respectively. It is worth noting that, during the train and augment process of DiffuASR and competing methods, we follow~\cite{kim2023melt} to exclude the test item of each sequence to avoid data leakage. For the stability of the results, all reported performances are averaged on three runs of SRS models with different random seeds. The implementation code is available online to ease reproducibility\footnote{\url{https://github.com/liuqidong07/DiffuASR}}\footnote{\url{https://gitee.com/mindspore/models/tree/master/research/recommend/DiffuASR}}.

\subsubsection{\textbf{Evaluation Metrics}}

    In this paper, we adopt \textit{Top-10 Hit Rate} (\textbf{HR@10}) and \textit{Top-10 Normalized Discounted Cumulative Gain} (\textbf{NDCG@10}) as the evaluation metrics, which have been used in related works~\cite{sun2019bert4rec,kang2018self,zhou2020s3}. HR evaluates the capacity to recommend correct items that appear in the Top-K list. NDCG can reflect the ranking ability of the recommendation models. Following~\cite{kang2018self,zhou2020s3}, we adopt the negative sampling strategy, \ie the ground-truth item paired with 100 randomly sampled negatives. All the reported results are averaged over all test users.

\begin{table*}[t]
\centering
\caption{The overall results of competing augmentation methods and DiffuASR on three datasets. Boldface refers to the highest score and underline indicates the best result of the baselines. ``\textbf{{\Large *}}'' indicates the statistically significant improvements (\ie two-sided t-test with $p<0.05$) over the best baseline.}
\resizebox{0.7\textwidth}{!}{
\begin{tabular}{cl|cc|cc|cc}
\toprule[1.5pt]
\multirow{2}{*}{Backbone} & \multirow{2}{*}{Augment Method} & \multicolumn{2}{c}{Yelp} & \multicolumn{2}{c}{Beauty} & \multicolumn{2}{c}{Steam} \\ \cline{3-8} 
 &  & NDCG@10 & HR@10 & NDCG@10 & HR@10 & NDCG@10 & HR@10 \\ 
 \midrule
 \midrule
\multirow{6}{*}{Bert4Rec} & - None & \underline{0.4899} & \underline{0.7702} & 0.2581 & \underline{0.4302} & \underline{0.6104} & \underline{0.8561} \\ \cmidrule{2-8} 
 & - Random & 0.4559 & 0.7464 & \underline{0.2634} & 0.4297 & 0.6022 & 0.8505 \\ 
 & - Random-Seq & 0.4703 & 0.7269 & 0.2509 & 0.4257 & 0.5771 & 0.8306 \\ 
 & - ASReP & 0.4714 & 0.7563 & 0.2513 & 0.4102 & 0.5765 & 0.8289 \\ \cmidrule{2-8} 
 & - DiffuASR (CG) & \textbf{0.5004}* & \textbf{0.7820}* & \textbf{0.2818}* & \textbf{0.4462}* & \textbf{0.6173}* & \textbf{0.8615}* \\ 
 & - DiffuASR (CF) & \textbf{0.5066}* & \textbf{0.7828}* & \textbf{0.2752}* & \textbf{0.4385}* & \textbf{0.6156}* & \textbf{0.8609}* \\ 
 \midrule
 \midrule
\multirow{6}{*}{SASRec} & - None & 0.4813 & 0.7590 & 0.2844 & 0.4343 & 0.5829 & 0.8323 \\ \cmidrule{2-8} 
 & - Random & \underline{0.4817} & 0.7590 & 0.2808 & 0.4293 & 0.5837 & 0.8314 \\ 
 & - Random-Seq & 0.4546 & 0.6862 & 0.2746 & 0.4204 & 0.5510 & 0.8055 \\ 
 & - ASReP & 0.4758 & \underline{0.7604} & \underline{0.2882} & \underline{0.4364} & \underline{0.5866} & \underline{0.8348} \\ \cmidrule{2-8} 
 & - DiffuASR (CG) & \textbf{0.4872}* & \textbf{0.7623}* & \textbf{0.2987}* & \textbf{0.4424}* & \textbf{0.5899}* & \textbf{0.8371}* \\ 
 & - DiffuASR (CF) & \textbf{0.4885}* & \textbf{0.7617} & \textbf{0.2952}* & \textbf{0.4380} & \textbf{0.5891}* & \textbf{0.8362} \\ 
 \midrule
 \midrule 
\multirow{6}{*}{S3Rec} & - None & 0.4832 & 0.7632 & 0.2908 & \underline{0.4372} & \underline{0.5755} & \underline{0.8245} \\ \cmidrule{2-8} 
 & - Random & 0.4862 & 0.7666 & \underline{0.2951} & 0.4367 & 0.5720 & 0.8218 \\ 
 & - Random-Seq & 0.4518 & 0.6983 & 0.2560 & 0.4042 & 0.5326 & 0.7833 \\ 
 & - ASReP & \underline{0.4864} & \underline{0.7681} & 0.2655 & 0.4055 & 0.5709 & 0.8220 \\ \cmidrule{2-8} 
 & - DiffuASR (CG) & \textbf{0.4994}* & \textbf{0.7710} & \textbf{0.3022}* & \textbf{0.4415} & 0.5752 & \textbf{0.8251} \\ 
 & - DiffuASR (CF) & \textbf{0.4996}* & \textbf{0.7727}* & \textbf{0.3022}* & \textbf{0.4454}* & \textbf{0.5802}* & \textbf{0.8294}* \\
 \bottomrule[1.5pt]
\end{tabular}
}
\label{tab:overall}
\vspace{-3mm}
\end{table*}

\subsection{Overall Performance Comparison (RQ1)}

To prove effectiveness and generality, we show the performance comparison between DiffuASR and competing baselines on three SRS models in Table~\ref{tab:overall}. \textit{- DiffuASR (CG)} and \textit{- DiffuASR (CF)} represents the DiffuASR equipped with classifier guide and classifier free strategy, respectively. We analyze results in the following parts.

Overall, we find that DiffuASR (CG) and DiffuASR (CF) can consistently outperform competing augmentation methods on all three SRS models, illustrating that our DiffuASR can generate higher-quality augmentation. Then, observing the two baselines of random sampling, they underperform in most situations. Random-Seq samples items from the raw sequence, making the augmentation more similar to the original sequence. However, for long-tail users, this method causes a large amount of repetition of a few items, which makes the SRS models inclined to recommend these items repeatedly and thus degrades the performance. By comparison, Random augmentation generates noisy items as augmentation, but can elevate the performance slightly under some conditions, especially on more sparse dataset. We think the reason is that it can alleviate the problems of sparsity and long-tail users. Compared with DiffuASR, ASReP cannot bring more gains to SRS models, because it is limited by the inferior augmentation quality.

By analysis of the different datasets, we find that SRS models can get more promotion via augmentation methods on the Beauty dataset while few gains on the Steam dataset. The reason is that Beauty owns larger sparsity and shorter interaction sequences than Steam. This phenomenon illustrates that data sparsity and long-tail user problems degrade the SRS model exactly. Comparing the results among different SRS models, we find that Bert4Rec can only benefit from our DiffuASR, which indicates that Bert4Rec is more prone to noisy augmentations. As for SASRec, ASReP can perform better than other baselines. We think the reason may be that the generation model has the same architecture as the recommendation model, \ie SASRec, which shares the same transitional patterns. In contrast, DiffuASR can outperform among various SRS models, which proves the generality of the proposed DiffuASR.

In summary, we can conclude that our DiffuASR is effective and general in promoting SRS models, which can be regarded as the response to research question \textbf{RQ1}.

\subsection{Analysis of Different User Groups (RQ2)}

\begin{figure*}[t]
\centering
\includegraphics[width=0.85\linewidth]{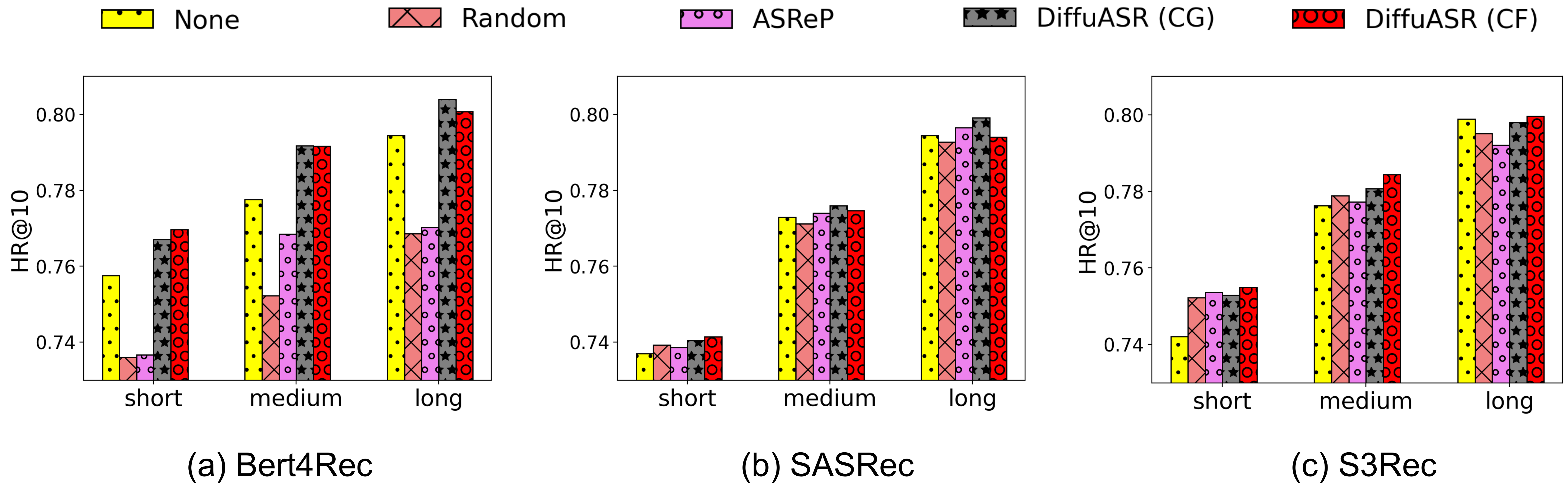}
\caption{The performance on Yelp dataset with grouped users. ``Short'' denotes the length of user's interaction in the range of [3, 5], ``Medium'' in the range of (5, 20], ``Long'' in the range of (20, $\infty$).}
\label{fig:exp_group}
\vspace{-2mm}
\end{figure*}

As mentioned above, the long-tail user problem is one challenge for SRS models. To answer the \textbf{RQ2}, we show the recommending performance on different groups of users with various numbers of historical interactions in Figure~\ref{fig:exp_group}. We divide the users into three groups based on their interaction number: \textit{short} ($3 \leq n_u \leq 5$), \textit{medium} ($5< n_u \leq 20$) and \textit{long} ($n_u>20$). The experimental results show that Bert4Rec has better recommending performance in the short group than the other two SRS models, indicating that Bert4Rec is less prone to the long-tail user problem. However, we can find that Random and ASReP degrade Bert4Rec on all three groups, because it is more sensitive to noisy augmentation. By comparison, DiffuASR elevates the performance due to the high-quality augmentation. As for SASRec and S3Rec, all augmentation methods promote short groups, which shows that they alleviate the long-tail user problem. Nevertheless, they show a performance decrease in long user group for noise reason. In contrast, the proposed DiffuASR degrades long user group less and even brings a little increase. In conclusion, the DiffuASR can better alleviate the long-tail user problem consistently and avoid too much damage for experienced users.


\begin{figure*}[!t]
\begin{minipage}[t]{0.245\linewidth}
\centering
\begin{subfigure}{1\linewidth}
    \includegraphics[scale=0.24]{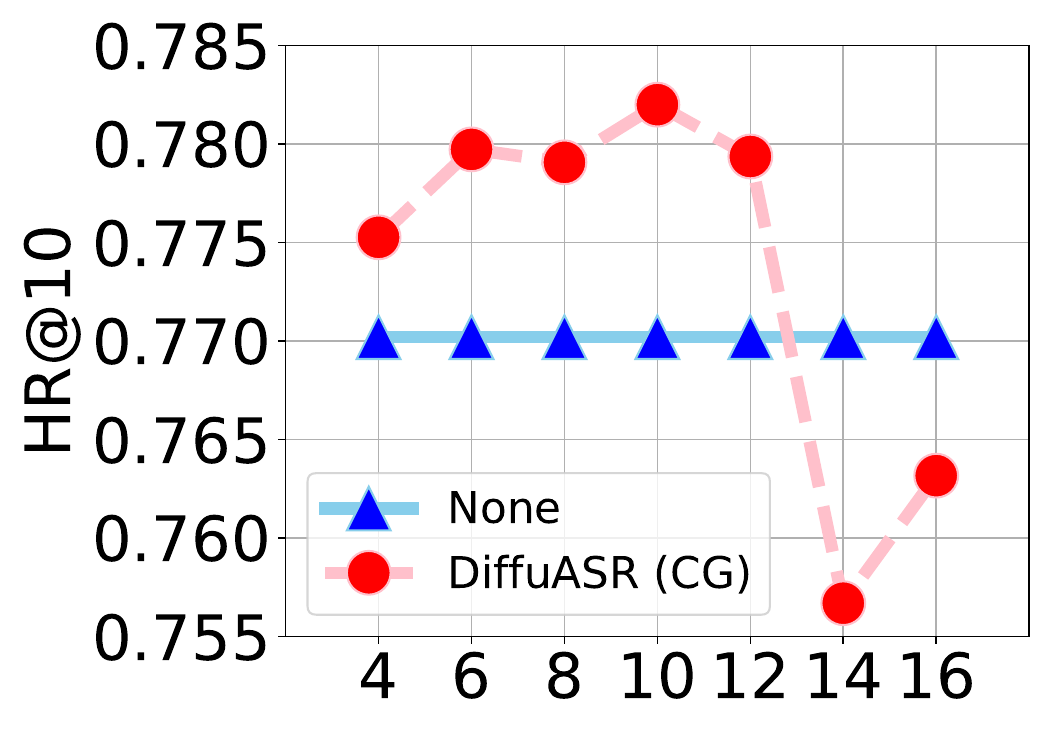}
    \caption{}
\label{fig:num_cg_hr}
\end{subfigure}
\end{minipage}%
\begin{minipage}[t]{0.245\linewidth}
    \centering
\begin{subfigure}{1\linewidth}
    \includegraphics[scale=0.24]{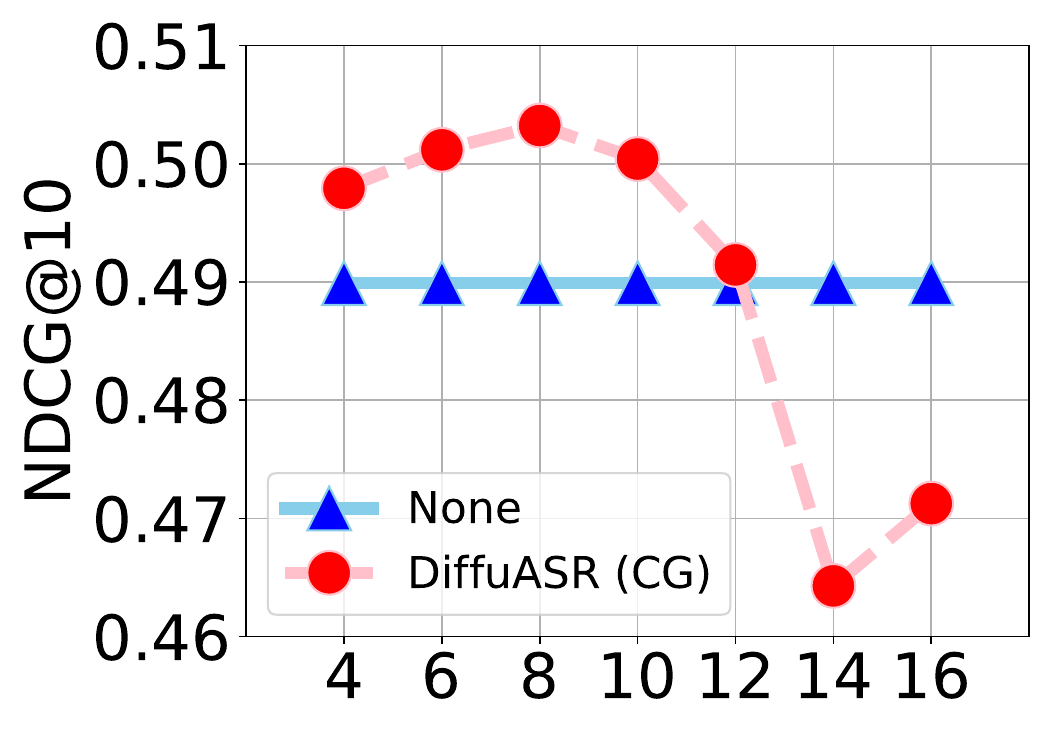}
    \caption{}
    \label{fig:num_cg_ndcg}
\end{subfigure}
\end{minipage}%
\begin{minipage}[t]{0.245\linewidth}
\centering
\begin{subfigure}{1\linewidth}
    \includegraphics[scale=0.24]{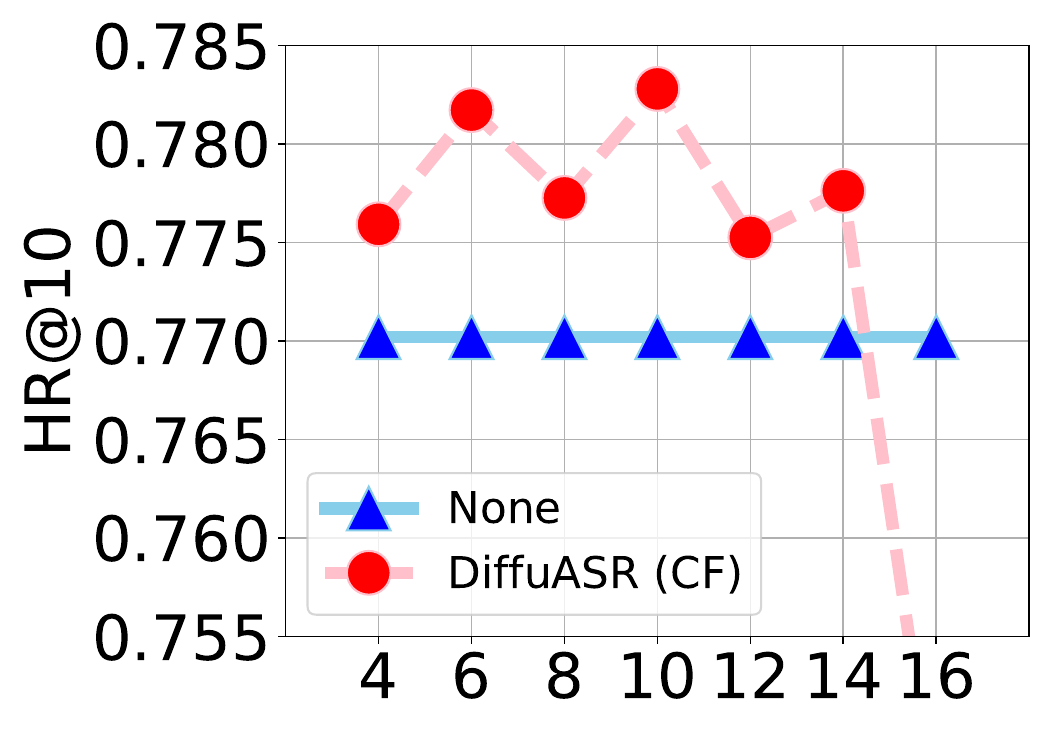}
    \caption{}
\label{fig:num_cf_hr}
\end{subfigure}
\end{minipage}
\begin{minipage}[t]{0.245\linewidth}
\centering
\begin{subfigure}{1\linewidth}
    \includegraphics[scale=0.24]{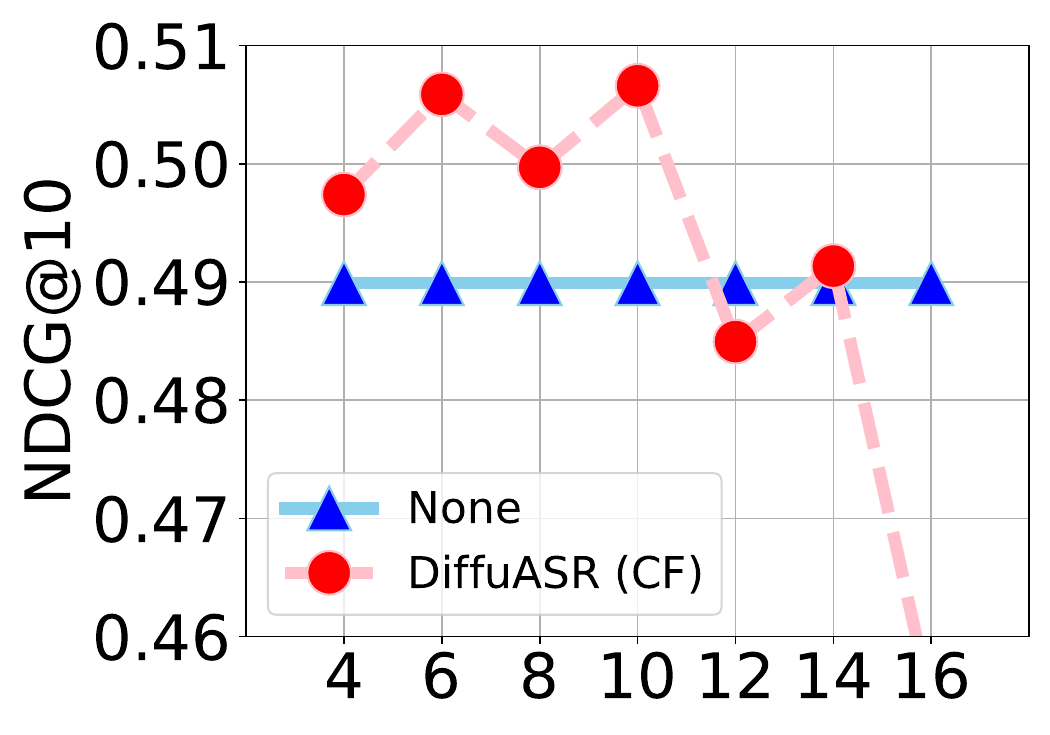}
    \caption{}
\label{fig:num_cf_ndcg}
\end{subfigure}
\end{minipage}
\caption{The results of experiments for the number of augmented items $M$ for each user. All the results are conducted on Yelp dataset and for Bert4Rec model.}
\label{fig:exp_num}
\end{figure*} 

\begin{figure*}[!t]
\vspace{-2mm}
\begin{minipage}[t]{0.245\linewidth}
\centering
\begin{subfigure}{1\linewidth}
    \includegraphics[scale=0.24]{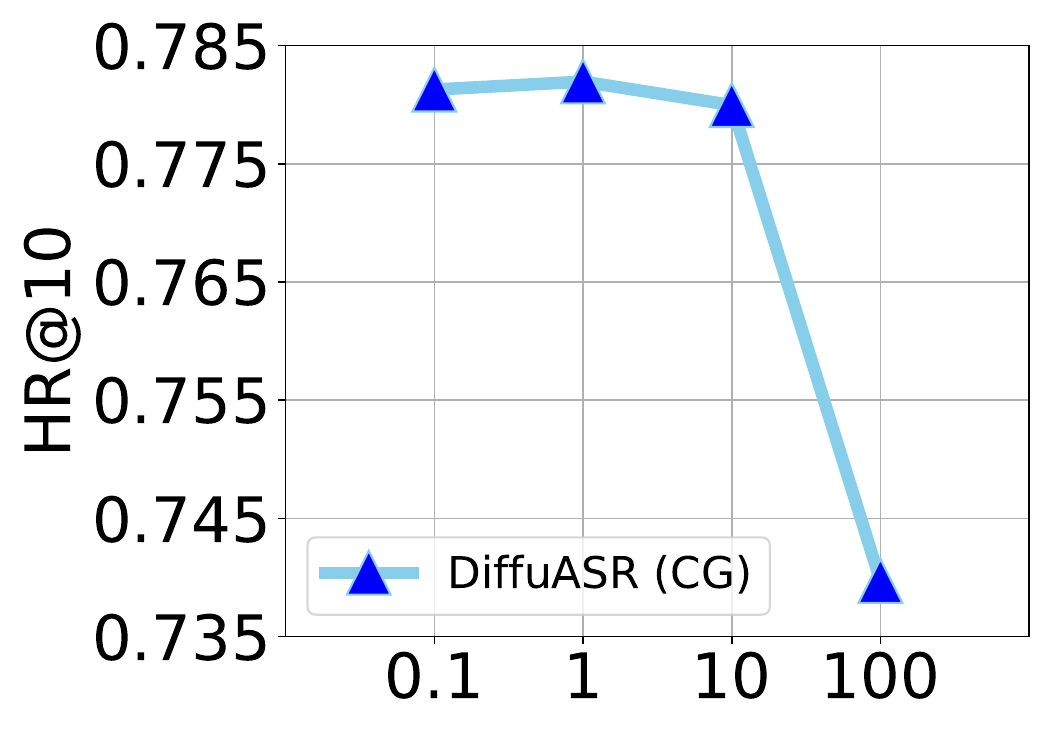}
    \caption{}
\label{fig:scale_cg_hr}
\end{subfigure}
\end{minipage}%
\begin{minipage}[t]{0.245\linewidth}
    \centering
\begin{subfigure}{1\linewidth}
    \includegraphics[scale=0.24]{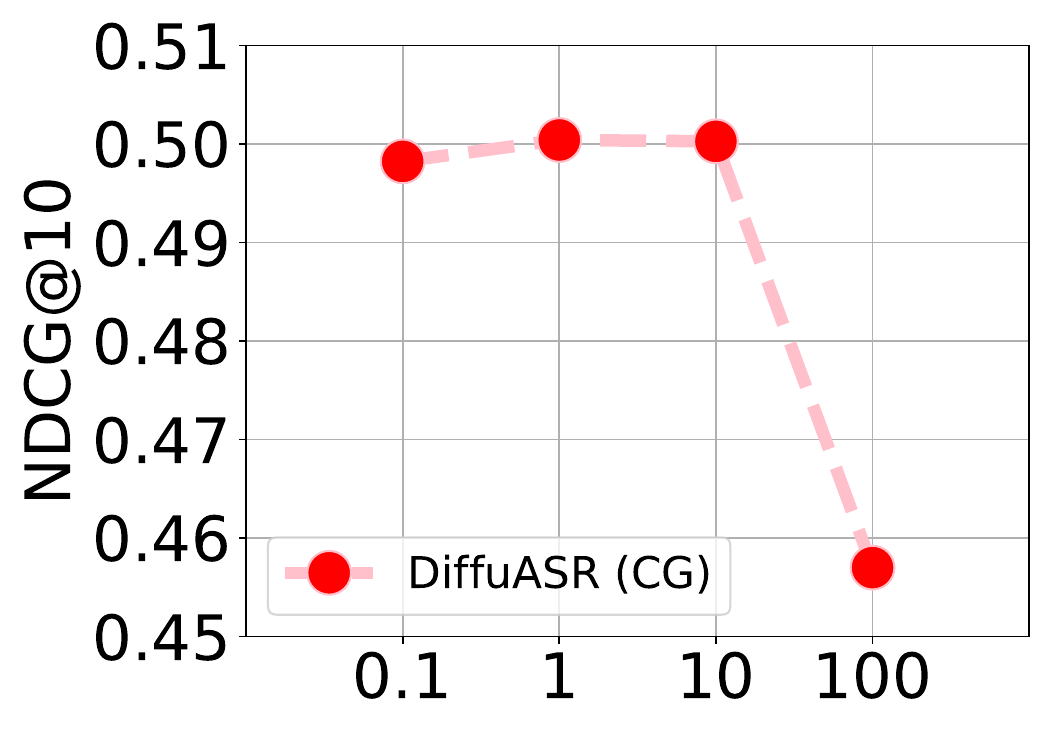}
    \caption{}
    \label{fig:scale_cg_ndcg}
\end{subfigure}
\end{minipage}%
\begin{minipage}[t]{0.245\linewidth}
\centering
\begin{subfigure}{1\linewidth}
    \includegraphics[scale=0.24]{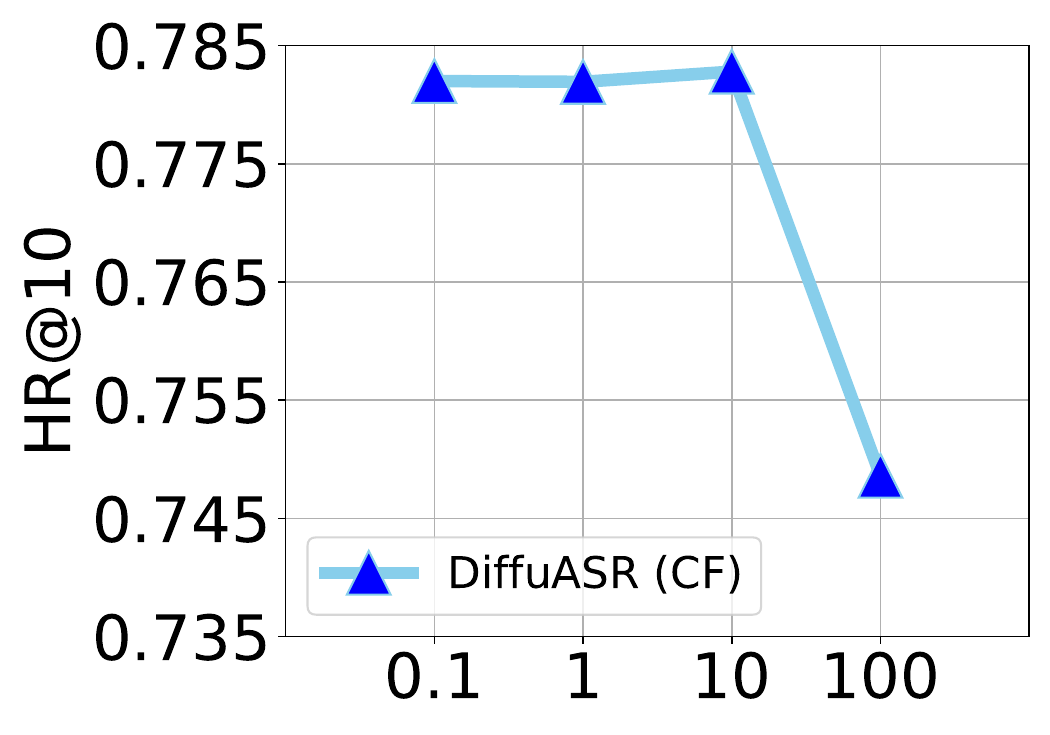}
    \caption{}
\label{fig:scale_cf_hr}
\end{subfigure}
\end{minipage}
\begin{minipage}[t]{0.245\linewidth}
\centering
\begin{subfigure}{1\linewidth}
    \includegraphics[scale=0.24]{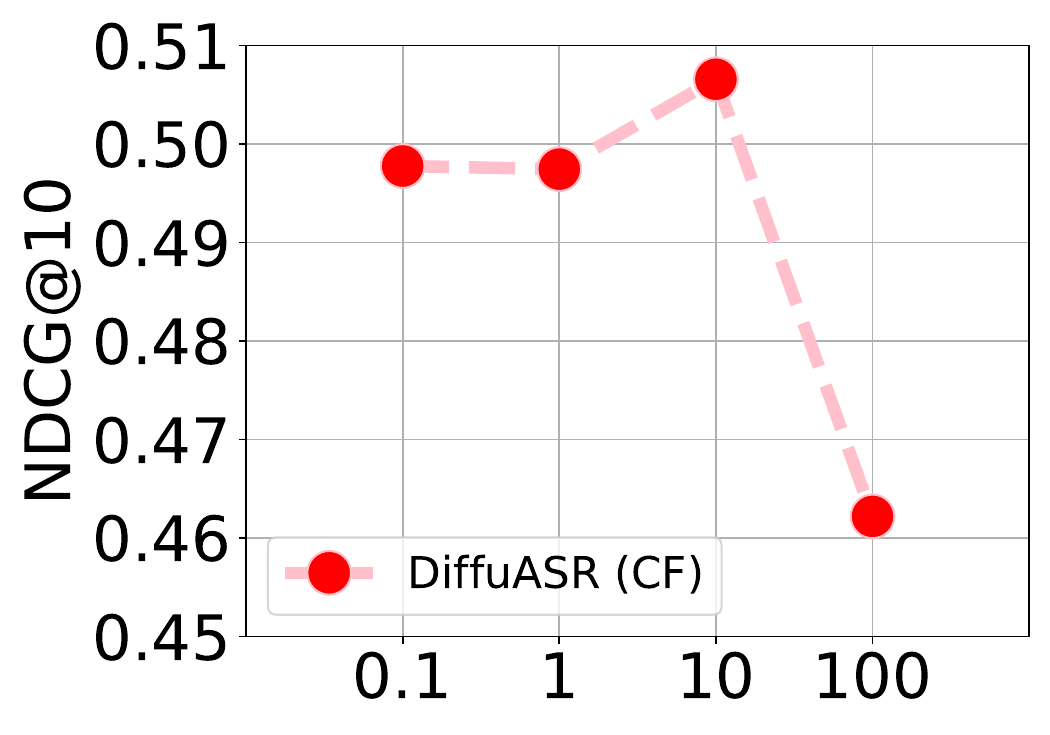}
    \caption{}
\label{fig:scale_cf_ndcg}
\end{subfigure}
\end{minipage}
\caption{The results of experiments for hyper-parameter $\gamma$. All the results are conducted on Yelp dataset and for Bert4Rec.}
\label{fig:exp_scale}
\vspace{-4mm}
\end{figure*} 

\subsection{Analysis of Augmented Number (RQ3)}

The augmentation can enrich the interactions for the whole dataset and each user, so the number of augmented items indicates to what extent the DiffuASR alleviates sparsity and long-tail user problem. For the response to the \textbf{RQ3}, we show the performance changes of Bert4Rec according to the augment number $M$ in Figure~\ref{fig:exp_num}. The experimental results show that the recommending performance gets increased when the augment number $M$ grows from 4 to 10, because more augment items can prolong the user's interaction sequence and reduce the sparsity of the whole dataset. However, we also find that if $M$ continuously rises from 10 to a larger number, the recommending performance drops and is even worse than no augmentation. The reason is that with the growth of $M$, the data that can be used for training DiffuASR becomes less. $\mathcal{D}_T$ only contains the interaction sequence with length $n_u<M$, because the first $M$ items are used for diffusion process in training process. With the decrease of training data for DiffuASR, the diffusion model easily falls into the underfitting problem, which leads to poor generation quality. In conclusion, the augmented number $M$ should take an appropriate value to maximize the benefits of data augmentation.

\subsection{Hyper-parameter Analysis (RQ4)}

As the response to \textbf{RQ4}, we explore the effects of two important hyper-parameters in DiffuASR, \ie schedule of $\beta$ and guidance scale $\gamma$. As illustrated in~\cite{nichol2021improved}, different schedules of $\beta$ fit for various data distributions, so we aim to explore the most suitable schedule for our DiffuASR. In this experiment, we test the linear~\cite{ho2020denoising} (\textit{- linear}), sqrt-root~\cite{li2022diffusion} (\textit{- sqrt}), cosine~\cite{nichol2021improved} (\textit{- cosine}) and sigmoid (\textit{- sigmoid}) schedules, whose results are shown in Table~\ref{tab:scheduler}. The experimental results indicate that the linear schedule performs the best for both classifier guide and classifier free variants. Therefore, we choose the linear schedule for $\beta$ across all the experiments in this paper. Besides, Dhariwa \etal~\cite{dhariwal2021diffusion} and Ho \etal~\cite{ho2022classifier} illustrate that the guidance scale controls the trade-off between fidelity and diversity of generation. To explore the effects of $\gamma$, we show the performance trends with the change of $\gamma$ in Figure~\ref{fig:exp_scale}. The experimental results show that when $\gamma$ increases in small values, the recommending performance of DiffuASR gets elevation, because the generated items are more corresponding to the intent contained in the raw sequence. However, when the $\gamma$ grows continuously, the performance of the SRS model drops drastically. We think it is because larger $\gamma$ leads to augmented items more similar to items in raw sequence, which causes a similar problem as the Random-Seq baseline.

\begin{table}[t]
\centering
\caption{The results on Yelp dataset and Bert4Rec model with different scheduler of $\beta$. The best results are in boldface.}
\begin{tabular}{lcccc}
    \toprule[1.5pt]
    \multirow{2}{*}{Scheduler} & \multicolumn{2}{c}{DiffuASR (CG)} & \multicolumn{2}{c}{DiffuASR (CF)} \\ 
    \cmidrule(lr){2-3} \cmidrule(lr){4-5}
     & NDCG@10 & HR@10 & NDCG@10 & HR@10 \\ 
    \midrule
    \midrule
    - linear & \textbf{0.5004} & \textbf{0.7820} & \textbf{0.5066} & \textbf{0.7828} \\
    - sqrt & 0.4440 & 0.7304 & 0.4495 & 0.7345 \\
    - cosine & 0.4510 & 0.7386 & 0.4779 & 0.7579 \\
    - sigmoid & 0.4512 & 0.7410 & 0.4533 & 0.7391  \\ 
    \bottomrule[1.5pt]
    \end{tabular}
\label{tab:scheduler}
\vspace{-6mm}
\end{table}

%% file: 5RelatedWork.tex
\section{Related Works}


\textbf{Sequential Recommendation}.
There are two main branches in the recommendation research field. One line aims to make full use of side information of users and items, so these works focus on feature interaction~\cite{rendle2010factorization,guo2017deepfm} and feature selection~\cite{wang2022autofield,lin2022adafs,wang2023single}. The other branch pays attention to modeling the user's historical records, \ie sequential recommendation, which predicts the next item based on the user's historical interactions. At the early stage, some traditional deep neural networks are adopted. For example, Hidasi \etal~\cite{hidasi2015session} and Tang \etal~\cite{tang2018personalized} propose to use RNN and CNN to extract preference from user's records, respectively. Recently, self-attention has shown the powerful ability for sequential modeling, which inspires researchers to adapt it to sequential recommendation, such as SASRec~\cite{kang2018self} and Bert4Rec~\cite{sun2019bert4rec}.
Besides, some works~\cite{zhou2020s3,li2022mlp4rec,li2023automlp,liang2023mmmlp} begin to merge the two branches mentioned above, \ie combining side information with the sequential recommendation. Some other works explore to alleviate the problem of popularity bias~\cite{liu2023disentangling,yang2023debiased}, efficiency~\cite{zhao2022mae4rec} and noise~\cite{zhang2022hierarchical,zhang2023denoising} for SRS. Despite the brilliant progress of SRS models, the data sparsity problem hinders further improvement. Two lines of works try to face this challenge, \ie data augmentation~\cite{zhang2021causerec,wang2021counterfactual,wang2022learning,chen2022data,liu2021augmenting,jiang2021sequential} and contrastive learning~\cite{xie2022contrastive,chen2022intent,qin2023meta}. 
We categorize data augmentation methods into two types: sequence-level and item-level. Sequence-level methods~\cite{zhang2021causerec,wang2021counterfactual,wang2022learning,chen2022data} often enrich the data by generating new sequences, such as L2Aug~\cite{wang2022learning} and CauseRec~\cite{zhang2021causerec}.
Though sequence-level methods can promote sequential recommendation models, they often need fabricated training procedures. Item-level methods~\cite{liu2021augmenting,jiang2021sequential} augment pre-order items for each sequence. 
ASReP~\cite{liu2021augmenting} firstly proposes to train a SASRec reversely as an augmentor and generate the pre-order items iteratively. Based on the augmentation of ASReP, BiCAT~\cite{jiang2021sequential} further adds a forward constraint and self-knowledge distillation to enhance the robustness of the model. ASReP and BiCAT pay more attention to how to use augmented datasets, but ignore poor augment quality, which may hinder the benefits of augmentation. Therefore, to further improve the augment quality, we propose a diffusion augmentation method.

\noindent\textbf{Diffusion Model}.
The diffusion model has attracted much attention recently, because of its superior performance on image generation. DPM~\cite{sohl2015deep} and DDPM~\cite{ho2020denoising} propose the general process of diffusion model. 
Since the reverse process is not parallel, the diffusion model suffers from low sampling speed. 
To accelerate the generation process, DDIM~\cite{song2021denoising} proposes to transform the forward process to a non-markovian process and thus can reduce the steps for the reverse process. Besides, control of generation is another vital issue. Dhariwal \etal~\cite{dhariwal2021diffusion} and Ho \etal~\cite{ho2022classifier} devise a classifier guide and classifier free diffusion model to control the generation corresponding to the desired class. ControlNet~\cite{zhang2023adding} proposes a novel finetuning strategy to add an additional control signal to a well-trained diffusion model. However, there are still few works to explore the applications of diffusion model in recommendation. Three recent works~\cite{wang2023diffusion,du2023sequential,li2023diffurec} take the first step. Wang \etal~\cite{wang2023diffusion} adapts the diffusion model to generative recommender system. DiffuRec~\cite{li2023diffurec} and DiffRec~\cite{du2023sequential} utilize the diffusion model for the sequential recommendation, but they generate the target item directly, which differs from the usage of DiffuASR. We use the generative ability of the diffusion model for data augmentation, which explores a novel way to adopt the diffusion model in the recommendation field.


%% file: 6Conclusion.tex
\section{Conclusion}

In this paper, we propose a diffusion augmentation for sequential recommendation (DiffuASR). To adapt the diffusion model to the item sequence generation, we first derive the framework of DiffuASR. Then, we design a Sequential U-Net to capture the sequence information while predicting the added noise. Besides, two guide strategies are designed to control the DiffuASR to generate the items more corresponding to the preference contained in the raw sequence. We conduct extensive experiments on three real-world datasets combined with three sequential recommendation models. The experimental results show the effectiveness and generality of our DiffuASR. Besides, we find that the number of augmented items affects performance largely in the experiments. 


%% file: 8Acknowledge.tex
\begin{acks}
    This research was partially supported by Huawei (Huawei Innovation Research Program), APRC - CityU New Research Initiatives (No.9610565, Start-up Grant for New Faculty of City University of Hong Kong), CityU - HKIDS Early Career Research Grant (No.9360163), Hong Kong ITC Innovation and Technology Fund Midstream Research Programme for Universities Project (No.ITS/034/22MS), SIRG - CityU Strategic Interdisciplinary Research Grant (No.7020046, No.7020074), SRG-Fd - CityU Strategic Research Grant (No.7005894), National Key Research and Development Program of China (2022ZD0117102), National Natural Science Foundation of China (No.62293551, No.62177038, No.62277042, No.62137002, No.61721002, No.61937001), Innovation Research Team of Ministry of Education (IRT\_17R86), Project of China Knowledge Centre for Engineering Science and Technology, Project of Chinese academy of engineering ``The Online and Offline Mixed Educational Service System for `The Belt and Road' Training in MOOC China''. We thank MindSpore \cite{mindspore} for the partial support of this work, which is a new deep learning computing framework.
\end{acks}